\documentclass[aps,article,nofootinbib,groupedaddress, twocolumn,superscriptaddress]{revtex4}
\usepackage{graphicx}
\usepackage{amsmath,amssymb}
\usepackage{hyperref}
\usepackage[usenames]{color}
\usepackage{url}
\usepackage{bm}

\hypersetup{
 colorlinks=true,
 linkcolor=red,
 citecolor=blue,
 urlcolor=black
}

\def\Journal#1#2#3#4{{#1} {\bf #2}, #3 (#4)}
\def\be{\begin{equation}}
\def\ee{\end{equation}}
\def\ba{\begin{eqnarray}}
\def\ea{\end{eqnarray}}

\def\f{\frac}
\def\l{\left}
\def\r{\right}

\def\om{\Omega_{\rm m}}
\def\XI{\mathcal{X}_{\rm I}}
\def\XII{\mathcal{X}_{\rm II}}
\def\Map{M_{\rm ap}}
\newcommand{\dd}{{\mathrm d}}
\newcommand{\J}{{\mathrm J}}

\newcommand\PRL{Phys.~Rev.~Lett.}

\begin{document}
\title{Probing modifications of General Relativity using current cosmological observations}
\date{\today}


\author{Gong-Bo Zhao}
\affiliation{Institute of Cosmology and Gravitation, University of Portsmouth, Dennis Sciama Building, Burnaby Road, Portsmouth, PO1 3FX, UK}

\author{Tommaso Giannantonio}
\affiliation{Argelander-Institut f\"ur Astronomie der Universit\"at Bonn, Auf dem H\"ugel 71, D-53121 Bonn, Germany}

\author{Levon Pogosian}
\affiliation{Department of Physics, Simon Fraser University,
Burnaby, BC, V5A 1S6, Canada}

\author{Alessandra Silvestri}
\affiliation{Kavli Institute for Astrophysics and Space Research, MIT, Cambridge, MA 02139, USA}

\author{David J. Bacon}
\affiliation{Institute of Cosmology and Gravitation, University of Portsmouth, Dennis Sciama Building, Burnaby Road, Portsmouth, PO1 3FX, UK}

\author{Kazuya Koyama}
\affiliation{Institute of Cosmology and Gravitation, University of Portsmouth, Dennis Sciama Building, Burnaby Road, Portsmouth, PO1 3FX, UK}

\author{Robert C. Nichol}
\affiliation{Institute of Cosmology and Gravitation, University of Portsmouth, Dennis Sciama Building, Burnaby Road, Portsmouth, PO1 3FX, UK}

\author{Yong-Seon Song}
\affiliation{Institute of Cosmology and Gravitation, University of Portsmouth, Dennis Sciama Building, Burnaby Road, Portsmouth, PO1 3FX, UK}

\begin{abstract}

{We test General Relativity (GR) using current cosmological data: the cosmic microwave
background (CMB) from WMAP5 (Komatsu {\it et al.} 2009), the
integrated Sachs-Wolfe (ISW) effect from the cross-correlation of
the CMB with six galaxy catalogs (Giannantonio {\it et al.} 2008), a compilation of
supernovae Type Ia (SNe) including the latest
SDSS SNe (Kessler {\it et al.} 2009), and part of the weak lensing (WL) data from CFHTLS (Fu
{\it et al.} 2008, Kilbinger {\it et al.} 2009) that probe linear and mildly non-linear scales.
We first test a model where the effective Newton's constant, $\mu$, and the ratio of the two
gravitational potentials, $\eta$, transit from the GR value to another
constant at late times; in this case, we find that GR is fully
consistent with the combined data. The strongest constraint comes from the ISW
effect which would arise from this gravitational transition; the
observed ISW signal imposes a tight constraint on a combination of
$\mu$ and $\eta$ that characterizes the lensing potential.
Next, we consider four pixels in time and space for each function
$\mu$ and $\eta$, and perform a Principal Component Analysis (PCA) finding that seven of the resulting eight eigenmodes are consistent with GR within the errors. Only one eigenmode shows a $2\sigma$ deviation from the GR prediction, which is likely to be due to a systematic effect. However, the detection of such a deviation demonstrates the power of our time- and scale-dependent PCA methodology when combining observations of structure formation and expansion history to test GR.}

\end{abstract}

\maketitle

\section{Introduction}

As cosmological observations improve, new possibilities arise for testing the physics that governs the evolution of our Universe. Precise all-sky measurements of the CMB by the WMAP satellite \cite{Larson:2010gs} have established that cosmic structure developed from a nearly scale-invariant initial spectrum of adiabatic fluctuations \cite{Komatsu:2010fb}. Baryon acoustic oscillations from SDSS \cite{Percival:2009xn} and growing catalogs of supernovae \cite{Kessler:2009ys}, in combination with the CMB, have tightened the constraints on the background expansion history, indicating a strong preference for the cosmological concordance model, $\Lambda$CDM. Correlating the CMB anisotropies from WMAP with wide-sky catalogs of galaxy counts have made it possible to detect the ISW effect \cite{Sachs:1967er, Crittenden:1995ak}, obtaining independent evidence for the accelerating expansion of the Universe \cite{Giannantonio:2008zi,Ho:2008bz}. Weak Lensing measurements by surveys such as COSMOS \cite{Massey:2007gh,Schrabback:2009ba} and CFHTLS \cite{Fu:2007qq,Kilbinger:2008gk} have afforded the use of shear correlation functions and power spectra in order to test cosmology.

In parallel, the problem of cosmic acceleration has motivated
explorations of new theoretical ideas, including the possibility
that GR may be modified on large scales.
Anticipating the substantial improvement in cosmological data
sets that is expected with surveys such as
DES \cite{DES}, Pan-STARRS \cite{Pan-STARRS}, LSST \cite{LSST} and Euclid \cite{Euclid}, model-independent
frameworks for testing GR against observations of the growth of cosmic
structures have been developed \cite{Linder:2007hg, Caldwell:2007cw, Zhang:2007nk, Amendola:2007rr, Hu:2007pj, Uzan:2006mf, Acquaviva:2008qp, Bertschinger:2008zb, MGCAMB, Pogosian:2010tj}.
Recently, there has been progress towards a
consensus on what properties such a framework should have
\cite{Hu:2007pj,MGCAMB,Pogosian:2010tj}.
First, it should be general, i.e. it should be able to describe a wide range of modified gravity models. Second, it
should not violate the consistency of super-horizon sized perturbations
with the background expansion. Third, it should involve as few
parameters as possible, while still being flexible enough to capture
most of the significant information contained in the data.

A good pragmatic starting point is to look for evidence of
departures in the fundamental relationships among the
perturbative fields familiar in cosmology: matter density and velocity
perturbations, and the metric perturbations \cite{ Zhang:2007nk,
Jain:2007yk, Song:2008vm, Song:2008xd}. It is commonly agreed that
one needs two general functions of scale and time, in addition to the conservation of energy-momentum, to specify the
evolution of the linear perturbations. For example, one can
introduce $\mu$, relating the gravitational potential to the density
contrast, and $\eta$, which relates the gravitational potential to
the spatial curvature. Given these functions, it is possible to
calculate the cosmological perturbations and all the observables
with a full Boltzmann integration code, such as {\tt MGCAMB}
\cite{MGCAMB,Pogosian:2010tj}.

The actual parametrization of the functions has not yet been
researched as fully. The goal is to strike a balance between
simplicity, i.e. working with as few parameters as possible, and
allowing for enough flexibility in these functions to capture all of
the information contained in the data. For example, one can
discretize $\mu(k,z)$ and $\eta(k,z)$ on a grid in $(k,z)$ space and
treat their values at each grid point, which we will
call {\it pixels}, as independent parameters \cite{Zhao:2009fn}.
Motivated by simplicity, one might wish to start with the simplest
possible model, such as constant values of $\mu$ and $\eta$, and
add complexity  only if the fitted parameter values of this simple
model show hints of departure from $\Lambda$CDM. However, this logic
may not work in modified gravity studies and might lead to missing
important information contained in the data. As the PCA analysis in
\cite{Zhao:2009fn} shows, 
the shapes of the well constrained eigenmodes suggest a higher sensitivity to scale-dependent features in $\mu$ and $\eta$, compared to their average values or the time dependence.

In this work, we first use a Fisher forecast-based PCA to determine the minimum number of pixels necessary to describe the shape of the best constrained eigenmodes of $\mu$ and $\eta$. We then use a Monte Carlo Markov Chain (MCMC) algorithm to fit these parameters to a combination of the available data, including CMB, ISW, SNe and WL, and find constraints on their de-correlated combinations. Throughout the paper we assume that the background expansion is given by the flat $\Lambda$CDM model which is strongly favored by the current constraints on the expansion history, and look for deviations from its predictions for the density perturbations. From a theoretical perspective, flatness is motivated by the inflationary origin of the Universe, and the viable models of modified gravity studied in the literature tend to be indistinguishable from $\Lambda$CDM at the background level.

Other tests of GR have been performed, in which different choices were made for the functions $\mu$ and $\eta$ (or a
related set of parameters). In \cite{Daniel,Daniel:2010ky} they were taken to have
a specific form of time dependence, while in  \cite{Bean:2010zq} they
have a specific form of time and scale dependence; finally,  
\cite{Daniel:2010ky} allowed them to vary in three redshift bins.
The results of these studies show a good consistency with $\Lambda$CDM.
To compare with
the results of some of these studies, we consider a single-transition-in-redshift model in
addition to our scale-dependent PCA method, generally finding a good
agreement with GR. In addition to allowing for scale-dependence, other important differences between our study and the treatment in \cite{Daniel:2010ky} include using the ISW cross-correlation data, using the CFHTLS WL data coming only from linear and mildly non-linear scales, and simultaneously varying two functions $\mu$ and $\eta$, while \cite{Daniel:2010ky} varied them one at a time when working with a $3$-bin model. The key conceptual difference from \cite{Bean:2010zq} is that we do not use WL data from a deeply non-linear regime, and we do not assume a specific scale and time dependence of the functions $\mu$ and $\eta$, but rather perform a  PCA of their values on a gird in $(k,z)$.

We find that the agreement with $\Lambda$CDM is
statistically more significant when no scale-dependence is allowed. After performing a PCA, we find that seven of the eight eigenmodes are consistent with GR within the errors. One eigenmode shows a $2\sigma$ deviation from the GR prediction, but can be directly traced to a feature in the WL aperture mass dispersion spectrum at $120$ arc min, which is most likely caused by a systematic \cite{wl_private}. However, the detection of this effect shows the benefits of adopting a more flexible scale-dependent pixellation of $\mu$ and $\eta$, and demonstrates that using scale-independent methods could potentially hinder the detection of new physics, or, as in this case, simply the better understanding of the data.

The paper is organized as follows.  Section~\ref{sec:data} reviews the data sets used in this work. After describing our
parametrization in Section~\ref{sec:para}, we present the constraints on departures from GR in Section~\ref{sec:results}, and finally draw
conclusions in Section~\ref{sec:concl}.

\section{Observables and Data} \label{sec:data}

In this section, we summarize the observables that will be used to
constrain deviations from GR, and explain the data sets for these
observables.

\subsection{Integrated Sachs-Wolfe effect}
The ISW effect \cite{Sachs:1967er} is a secondary anisotropy of the CMB which is created whenever the gravitational potentials are evolving in time. This is due to the net energy gain that the CMB photons acquire when traveling through varying potential wells, and it is therefore a direct probe of the derivatives of the potentials $\Phi, \Psi$.
In more detail, this effect generates additional CMB temperature anisotropies in any direction $\hat {\bf{n}}$ given by
\be
\Theta_{\mathrm{ISW}}  (\hat {\bf{n}}) \equiv \frac {\Delta T_{\mathrm{ISW}}} {T_{\mathrm{ISW}}} (\hat {\bf{n}}) = - \int \left( \dot \Phi + \dot \Psi \right) \left[ \tau, \hat {\bf{n}} (\tau_0 - \tau)  \right] d \tau,
\ee
where $ \tau $ is the conformal time, the dot represents a conformal
time derivative and the integral is calculated along the line of sight of
the photon.

A direct measurement of this effect is difficult, due to the overlap with the primary CMB anisotropies, whose amplitude is at least 10 times bigger. An additional problem is that the ISW signal is biggest on the largest angular scales, which are most affected by cosmic variance.
It is nevertheless possible to detect this signal by cross-correlating the full CMB with some tracers of the large-scale structure (LSS) of the Universe \cite{Crittenden:1995ak}: the primary CMB signal, generated at early times, is expected to have null correlation with the LSS, while the ISW anisotropies, produced at low redshift, correlate with the LSS distributions since they trace the fluctuations in the potentials.

We can then use a galaxy survey with visibility function $dN / dz (z)$ as a tracer of the LSS, and we can write the galaxy density fluctuation in a direction $\hat {\bf {n}}_1$ as
\be
\delta_g (\hat {\bf {n}}_1) =  \int b_g (z) \frac{dN}{dz}(z) \, \delta_m (\hat {\bf {n}}_1, z) \, dz,
\ee
where $b_g$ is the galaxy bias and $\delta_m$ the matter density perturbation. On the other hand, the ISW temperature anisotropy in a direction $\hat {\bf {n}}_2$  is given by
\be
\Theta_{\mathrm{ISW}}  (\hat {\bf {n}}_2) = - \int e^{-\kappa(z)} \frac {d } {dz} \left( \Phi + \Psi \right) (\hat {\bf {n}}_2, z) dz,
\ee
where $ e^{-\kappa(z)} $ is the photons' visibility function and $\kappa$ the optical depth. After choosing a particular data set for the CMB and the LSS, we can then define the auto- and cross-correlation functions as
\ba
c^{Tg}(\vartheta) & \equiv & \langle \Theta (\hat {\bf {n}}_1) \, \delta_g (\hat {\bf {n}}_2)  \rangle \\
c^{gg}(\vartheta) & \equiv & \langle \delta_g (\hat {\bf {n}}_1) \, \delta_g (\hat {\bf {n}}_2)  \rangle,
\ea
where $\Theta$ is the full CMB temperature anisotropy and the averages are calculated over all pairs at an angular
separation $ \vartheta = | \hat {\bf {n}}_1 - \hat {\bf {n}}_2 |$.
Alternatively, the above calculation can be written in harmonic space, and the auto- and cross-power spectra are then derived.

We use the ISW data from \cite{Giannantonio:2008zi}, which were obtained by cross-correlating multiple galaxy catalogs with the CMB maps from WMAP. The data used trace the distribution of the LSS in various bands of the electromagnetic spectrum, with median redshifts $0.1 < \bar z < 1.5$, and consist of six catalogs (infrared 2MASS, visible SDSS main galaxies, luminous red galaxies and quasars, radio NVSS, and X-ray HEAO). This is an approximation of a true tomographic study of the ISW signal.

All maps were pixellated on the sphere, with a pixel size of 0.9 deg.
The measurements were done in real space, calculating the angular
cross-correlation functions (CCFs) between the maps. These were
linearly binned in steps of 1 deg for angles $0$ deg $ \le \vartheta
\le 12$ deg; so the data set consists of 78 points $(c_i^{Tg})_{\rm obs}$.

A well known property of the correlation functions is that their data points are highly correlated; in this case, in particular, the high degree of correlation is present also between data points belonging to different catalogs, due to the partial overlaps in redshift and in sky coverage of the sources. For this reason, the full covariance matrix between all data points ${\bf {C}}_{ij}$ is a very important piece of information, and it was estimated in \cite{Giannantonio:2008zi} using several Monte Carlo and jack-knife methods.
Here we use the matrix produced with the most complete technique, a full Monte Carlo
method where both galaxies and CMB maps were simulated and then
correlated to measure the expected noise and covariance.

The calculation of the likelihood of a particular model given the ISW
data is done as follows. First, the theoretical CCFs $(c_i^{Tg})_{\rm theo}$ and
auto-correlation functions (ACFs) are calculated with a full
Boltzmann integration within {\tt MGCAMB}, based on the redshift
distributions of the sources. The galaxy bias parameters are assumed
to be independent and constant for each catalog, and are rescaled
for each model imposing that the ACFs on small angular scales match
the observations. Finally, the theoretical CCFs are multiplied by
this rescaled bias to calculate the $\chi^2_{\mathrm{ISW}}$
distribution, given by
\ba
\chi^2_{\mathrm{ISW}} &=& \sum_{ij}  \left[
(c_i^{Tg})_{\rm obs} - (c_i^{Tg})_{\rm theo} \right] \left[ {\bf {C}}^{-1} \right]_{ij}  \nonumber \\
&~& \times \left[ ( c_j^{Tg})_{\rm obs} - (c_j^{Tg})_{\rm theo} \right].
\ea

\subsection{Supernovae and Cosmic Microwave Background}

For the SNe data, we use the sample combination labeled $(e)$ shown in table 4
of \cite{Kessler:2009ys}, which is a compilation of the SDSS-II
SNe sample plus Nearby SNe, ESSENCE, SNLS and HST. To calculate
the SNe likelihood, we use values from the {\tt MLCS2K2} light curve fitter, and marginalize
over the nuisance parameter, which is the calibration uncertainty in
measuring the supernova intrinsic magnitude. Note that
\cite{Kessler:2009ys} found a discrepancy of the constraints on
the F$w$CDM model (standard CDM model in a flat Universe plus a dark
energy component with a constant $w$) using the two SNe fitters
{\tt MLCS2K2} and {\tt SALT-II}. However, the discrepancy is much smaller (within $1 \sigma$) for a
flat $\Lambda$CDM model as shown in Tables 13 and 17 in
\cite{Kessler:2009ys}, namely, \ba&&\om=0.312\pm0.022({\rm
stat})\pm0.001({\rm syst})~~({\tt MLCS2K2})\nonumber\\
&&\om=0.279\pm0.019({\rm stat})\pm0.017({\rm syst})~~({\tt
SALT-II})\nonumber\ea Since we will assume the background evolution is
the same as that in the flat $\Lambda$CDM model, the choice of the
SNe fitter does not affect our final results significantly, but
we should bear in mind that systematic errors are now comparable
to statistical errors in SNe observations.

For the Cosmic Microwave Background data in our analysis, we use the
WMAP five-year data including the temperature and polarization power
spectra~\cite{Komatsu:2008hk,WMAP5}, and calculate the likelihood
using the routine supplied by the WMAP
team\footnote{\url{http://lambda.gsfc.nasa.gov/}}.

\subsection{Weak lensing}\label{sec:WL}

We use the cosmic shear observations from the CFHTLS-Wide third year
data release T0003~\cite{Fu:2007qq, CFHTLS}, in which about $2\times 10^6$
galaxies with $i_{\rm AB}$-magnitudes between $21.5$ and $24.5$ were
imaged on 57 sq. deg. (35 sq. deg. effective area).
We use the aperture-mass dispersion
$\Map$~\cite{Schneider:1997ge} following
\cite{Fu:2007qq, Kilbinger:2008gk}. As in those studies, one can obtain the relevant $\chi^2$
by fitting the theory-predicted aperture-mass dispersion $\langle
M_{\rm ap}^2 \rangle_{\rm theo}$ given by a model parameter vector
$\mathbf{p}$ to $\langle M_{\rm ap}^2 \rangle_{\rm obs}$ measured at
angular scales $\theta_i$,
\ba
\chi^2_{\rm Map}(\mathbf{p}) &=& \sum_{ij}
\Big(\big \langle M_{\rm ap}^2(\theta_i)
\big\rangle_{\rm obs} -  \big\langle M_{\rm ap}^2(\theta_i, \mathbf{p})
\big\rangle_{\rm theo}
\Big) [\mathbf{C}^{-1}]_{ij} \nonumber \\
& & \times
\Big(\big \langle M_{\rm ap}^2(\theta_j)
\big\rangle_{\rm obs} -  \big\langle M_{\rm ap}^2(\theta_j, \mathbf{p})
\big\rangle_{\rm theo}
\Big).
 \label{chi2_lens}
\ea
Note that the data covariance matrix  $\mathbf{C}$
is the one used in \cite{Fu:2007qq} and \cite{Kilbinger:2008gk}; it contains shape
noise, (non-Gaussian) cosmic variance and residual B-modes \cite{wl_private}.
Since
it is difficult to model the weak lensing nonlinearity
in modified gravity in a model-independent way~\cite{Beynon:2009yd},
we only use the aperture-mass dispersion data measured between $30$
and $230$ arc min, to remove the strongly nonlinear region from the data.
For angles smaller than $30'$, the difference between  linear and non-linear predictions becomes greater than a factor of two, and we do not wish to suppose that non-linear corrections are reliable on smaller scales.

Theoretically, the aperture-mass dispersion is related to the weak
lensing power spectrum via~\cite{Schneider:1997ge}
\be\label{eq:Map}
\langle M_{\rm ap}^2 \rangle(\theta) = \int \frac{\dd \ell\,
 \ell}{2\pi} P_\kappa(\ell) \left[ \frac{24 \, \J_4(\theta
 \ell)}{(\theta \ell)^2}\right]^2 \, ,
\ee
for the choice of filter in \cite{Fu:2007qq}; here,  the lensing power spectrum $P_\kappa$ is a projection of the
3D matter-density power spectrum $P_\delta$, weighted by the
source galaxy redshift distribution and geometric factors, and ${\rm J}_{\alpha}(x)$ is the Bessel
function of the first kind. To model the redshift distribution of
the galaxies, we follow \cite{Fu:2007qq} and use the
parametrization
\be
n(z) \propto \frac{z^a + z^{ab}}{z^b + c} ; \;\;\;\; \int_0^{z_{\rm
   max}} n(z) \, \dd z = 1 \, ,
\label{eq:nz}
\ee
where $\mathcal{N}\supset\{a, b,c\}$ is a set of nuisance
parameters to be marginalized over, and we have imposed Gaussian priors
on them following \cite{Fu:2007qq}:
$a=0.612\pm0.043,~b=8.125\pm0.871,~c=0.620\pm0.065$. The distribution is normalized by setting
$z_{\rm max} = 6$.
Then we can calculate the
$\chi^2$ for the redshift uncertainty as,
\be
\chi_{z}^2 =  \sum_i \frac{[n_i -
   n(z_i)]^2}{\sigma_i^2}.
\label{eq:chi2nz}
\ee
where $n_i$ is the normalized number of galaxies in the $i^{\rm th}$
redshift bin and $n(z_i)$ the fitting function, evaluated at the
center of the redshift bin. As described in \cite{Fu:2007qq}, the
uncertainty $\sigma_i$ of $n_i$ contains Poisson noise, photo-$z$
error and cosmic variance, and we neglect the cross-correlation
between different bins. Then we obtain the $\chi^2$ for weak lensing
in the same way as \cite{Fu:2007qq}, \be \chi_{\rm WL}^2=
\chi^2_{\rm Map}+\chi_{z}^2 \, . \ee

\subsection{Further priors}
Finally, we impose  $1\sigma$ Gaussian priors on the Hubble
parameter and baryon density of $h=0.742\pm0.036$ and
$\Omega_{b}h^{2}=0.022\pm0.002$ from the measurements of Hubble
Space Telescope (HST) \cite{Riess:2009pu} and Big Bang
Nucleosynthesis \cite{BBN} respectively, and a top hat prior on the
cosmic age of 10 Gyr $< t_0 <$ 20 Gyr. The total likelihood is taken
to be the product of the separate likelihoods $\mathcal{L}$ of each
dataset we used; thus the total $\chi^2$ is the sum of separate
$\chi^2$ from individual observations plus that from the priors if
we define $\chi^2 \equiv -2 \log {\mathcal{L}}$.

\section{The parametrization of modified growth} \label{sec:para}

In order to test gravity against the growth of cosmological perturbations in a model-independent way, one needs a generalized set of equations to evolve linear perturbations without assuming GR.
We work within the framework of the Boltzmann integrator {\tt MGCAMB} \cite{MGCAMB}~\footnote{\url{http://userweb.port.ac.uk/
~zhaog/MGCAMB.html}}, which is a variant of {\tt CAMB} \cite{Lewis:1999bs}~\footnote{\url{http://camb.info/}}. This is based on a system of
equations that allows for a general modification of gravity at
linear order in the perturbations, while respecting the consistency
of the dynamics of long-wavelength perturbations with the background
expansion \cite{Wands:2000dp,Bertschinger:2006aw}. An interested
reader can find a detailed discussion of the equations used in {\tt
MGCAMB} and their comparison with other methods in the literature in
\cite{Pogosian:2010tj}.

We consider scalar metric perturbations about a FRW background for which the line element in the conformal Newtonian gauge reads
\be\label{metric}
ds^2=-a^2(\tau)\l[\l(1+2\Psi\r)d\tau^2-\l(1-2\Phi\r) d\vec{x}^2\r] \ ,
\ee
where $\Phi$ and $\Psi$ are functions of time and space. We assume adiabatic initial conditions and covariant conservation of the energy-momentum tensor of matter. The matter conservation at linear order provides two equations which in Fourier space can be written as
\begin{eqnarray}
\label{matter-conservation}
\delta'+{k\over aH}v-3\Phi'&=&0, \\
v'+v-{k\over aH}\Psi&=&0 \ ,
\label{matter-continuity}
\end{eqnarray}
where $\delta$ is the energy density contrast, $v$ the irrotational
component of the peculiar velocity, and primes indicate derivatives
with respect to  $\ln a$. In order to solve for the evolution of the
four scalar perturbations $\{\delta,v, \Phi,\Psi\}$, we need two
additional equations, provided by a theory of gravity (such as GR),
which specify how the metric perturbations relate to each other, and
how they are sourced by the perturbations of the energy-momentum
tensor. One can parametrize these relations as
\begin{eqnarray}
\label{gamma}
\frac{\Phi}{\Psi}&=&\eta(a,k), \\
\label{parametrization-Poisson}
k^2\Psi&=&-4\pi G a^2 \mu(a,k) \rho\Delta \ ,
\end{eqnarray}
where $\Delta$ is the gauge-invariant comoving density contrast defined as
\begin{equation}
\Delta \equiv \delta + 3\f{aH}{k} v\, ;
\label{Def:Delta}
\end{equation}
$\eta(a,k)=\mu(a,k)=1$ in GR, while in an alternative model $\mu$ and $\eta$ can
in general be functions of both time and scale~\cite{Bertschinger:2008zb,MGCAMB,Afshordi:2008rd}.

Defining $\mu$ and $\eta$ in this particular way makes Eqs.~(\ref{matter-conservation})-(\ref{parametrization-Poisson}) consistent on all linear scales. As shown in \cite{Pogosian:2010tj}, on super-horizon scales $\mu$ naturally becomes irrelevant and we are left with only one function, $\eta$, as expected from the super-horizon consistency conditions \cite{Wands:2000dp,Bertschinger:2006aw}. Also, having $\mu$ defined through the Poisson equation involving $\Delta$, as opposed to $\delta$, allows for $\mu$ to be equal to unity on all scales for GR.

From $\{ \eta$, $\mu \}$, we can derive other parameters which may be more suitable for interpreting observational constraints. For example, since we are
using WL and ISW observations in this paper, we will be essentially measuring the
power spectra of the lensing potential $(\Phi+\Psi)$ and its time
derivative. On the other hand, $\eta$ is not probed directly by any
observable and would be highly degenerate with $\mu$, as also pointed
out by \cite{Daniel:2010ky}. For that reason, in addition to $\{ \eta$, $\mu \}$, we also present our results in terms of another function, $\Sigma$, defined as
\be\label{eq:Sigma}
\Sigma(a,k)\equiv-\frac{k^2(\Psi+\Phi)}{8\pi{G}\rho{a}^2\Delta}=\frac{\mu(1+\eta)}{2}\, .
\ee
Note that specifying $\mu$ and $\Sigma$ is equivalent to working with $\mu$ and $\eta$. We use both parametrizations to discuss the physics and to interpret our
final results.

Since we are interested in testing GR at late times, we assume
$\mu(a,k)=\eta(a,k)=1$ at early times, namely, for $z>z_s$ where $z_s$
denotes the threshold redshift. This is natural in the existing models of modified gravity that aim to explain the late-time acceleration, where departures from GR occur at around the present day horizon scale. Also, the success in explaining the BBN and CMB physics relies on GR being valid at high redshifts.

One could assume a functional parametrization
for $\mu$ and $\eta$, either motivated by a modified growth (MG) theory or by simplicity,
and fit the parameters to the
data~\cite{MGCAMB,Giannantonio:2009gi,Daniel:2010ky}. We adopt a different approach, and pixelize
$\mu(a,k)$ and $\eta(a,k)$ on a grid in time and scale, fitting their values
in each grid point to the data. We then solve the eigenvalue problem for the covariance of the pixels (i.e. perform a PCA) to find their independent linear combinations that can be compared with their prediction in GR~\cite{Zhao:2009fn}. As we will elaborate
later, the PCA method has several advantages, such as being
model-independent and degeneracy-free, although it is much more
computationally expensive. In this work, we will utilize both
functional fit and PCA strategies in order to search for any imprint of
modified gravity.

To begin with, we parametrize our Universe using: \be
\label{eq:paratriz} {\bf P} \equiv (\omega_{b}, \omega_{c},
\Theta_{s}, \tau, n_s, A_s, \mathcal{N}, \mathcal{X}) \ee where
$\omega_{b}\equiv\Omega_{b}h^{2}$ and
$\omega_{c}\equiv\Omega_{c}h^{2}$ are the physical baryon and cold
dark matter densities relative to the critical density respectively,
$\Theta_{s}$ is the ratio (multiplied by 100) of the sound
horizon to the angular diameter distance at decoupling, $\tau$
denotes the optical depth to re-ionization, and $n_s, A_s$ are the
primordial power spectrum index and amplitude, respectively. We also vary and marginalize over
several nuisance parameters denoted by $\mathcal{N}$ when performing the
likelihood analysis for weak lensing and SNe, as we will elaborate
later. The modification of gravity is encoded in $\mathcal{X}$, and
we consider two different kinds of MG parametrizations, $\mathcal{X}_I$ and $\mathcal{X}_{II}$, as described in the following subsections.
Finally, we assume a flat Universe and an effective dark energy equation of state $w=-1$ throughout the expansion history.

\begin{figure*} [htp]
\begin{center}
\includegraphics[scale=0.85]{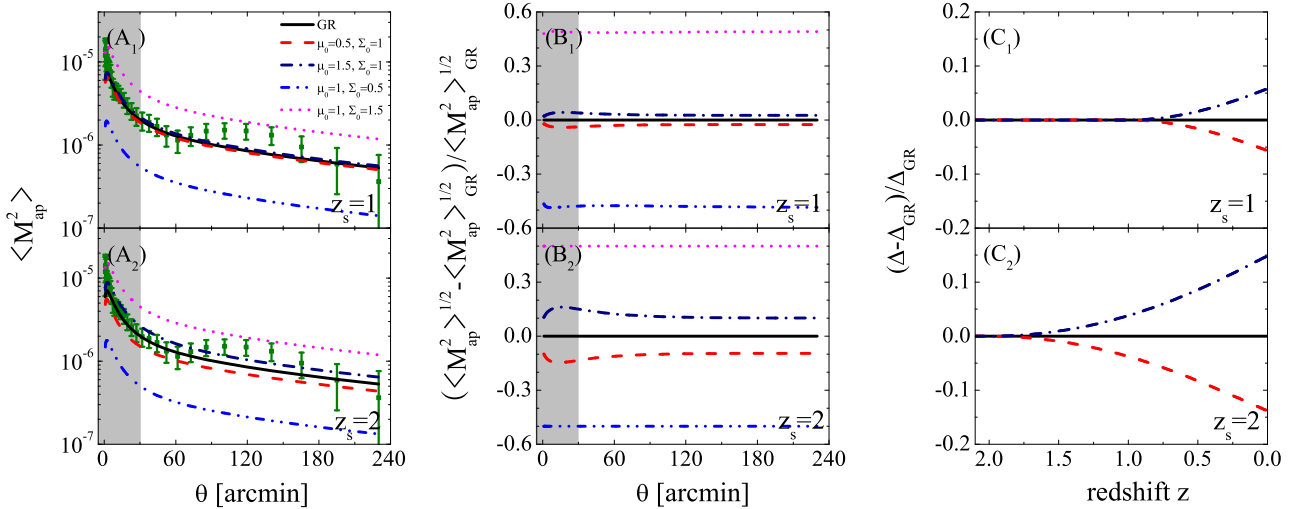}
\caption{Imprints of modified gravity parametrized by
$\mathcal{X}_{\rm I}$ on the weak lensing aperture mass dispersion
(panels $A_1, A_2$), relative difference in $\Map^{1/2}$ with
respect to GR ($B_1, B_2$), and relative difference in $\Delta$ with
respect to GR ($C_1, C_2$). The model parameters are shown in the
legend of panel ($A_1$). The shaded regions in panels
($A_1,A_2,B_1,B_2$) are excluded from our analysis. The data with
error bars over-plotted in panels ($A_1, A_2$) are taken from the
CFHTLS survey.}
\label{fig:WL_imprint}
\end{center}
\end{figure*}

\begin{figure} [htp]
\begin{center}
\includegraphics[scale=0.7]{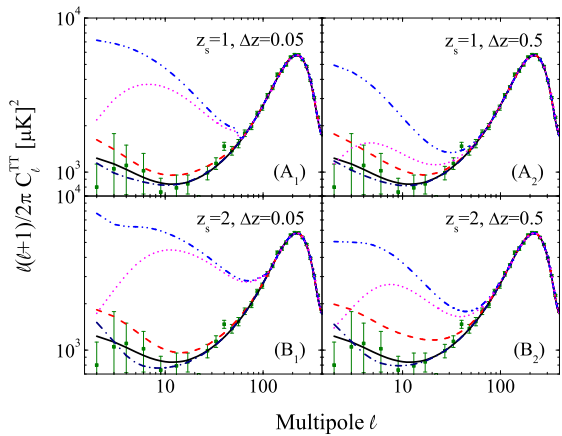}
\caption{Imprints of modified gravity parametrized by
$\mathcal{X}_{\rm I}$ on the CMB TT power spectra for different
threshold redshift $z_s$ and different transition width $\Delta{z}$.
Different models are distinguished by different line styles and
colors, as shown in panel ($A_1$) of Fig.~\ref{fig:WL_imprint}. The
data points with error bars are taken from the WMAP 5-year survey.}
\label{fig:CMB_imprint}
\end{center}
\end{figure}

\subsection{A single high redshift transition in $\mu$ and $\eta$: $\mathcal{X}_{\rm I}=\{\mu_0, \eta_0,
\Delta{z}\}$}\label{sec:XI}

There is no physical reason to assume that departures from GR ought to be scale-independent,
and a PCA forecast analysis~\cite{Zhao:2009fn} actually showed that the scale-dependence of $\mu$ and $\eta$ is
better constrained than their average values or the time-dependence.
Nevertheless, we shall first consider the case
in which $\mu$ and $\eta$ are taken to be scale-independent and transit from their GR values to another
constant value below a threshold redshift $z_s$.  Aside from simplicity, this will allow an insightful comparison with the results of the scale-dependent analysis later.

To model
the time evolution of $\mu$ and $\eta$ we use the hyperbolic
tangent function to describe the transition from unity to the constants
$\mu_0$ and $\eta_0$ as
\ba \mu(z)&=&\frac{1-\mu_0}{2}\Big(1+{\rm
tanh}\frac{z-z_s}{\Delta{z}}\Big)+\mu_0~,\nonumber\\
\eta(z)&=&\frac{1-\eta_0}{2}\Big(1+{\rm
tanh}\frac{z-z_s}{\Delta{z}}\Big)+\eta_0~. \ea
For a given $z_s$, the above parametrization has three free
parameters: $\mu_0, \eta_0$ and $\Delta{z}$. In
Figs.~\ref{fig:WL_imprint} and~\ref{fig:CMB_imprint}, we show the
imprints of MG on the weak lensing aperture-mass dispersion (see details
in Sec.~\ref{sec:WL}) and CMB TT power spectra for different values
of the MG parameters respectively. To be more physically
transparent, we will also present our results in terms of the $\{\mu,~\Sigma\}$ parametrization (see
Eq.~(\ref{eq:Sigma})).

We find that
both WL and CMB observables are more sensitive to the variation of
$\Sigma_0$ than to that of $\mu_0$ for both the $z_s=1$ and the $z_s=2$ case.
This is expected since WL and CMB (via the ISW effect) measure, respectively,  the power
spectra of $(\Phi+\Psi)$ and their time derivative, and are primarily controlled by $\Sigma_0$. However, at late times on sub-horizon
scales, $\mu_0$ only affects $(\Phi+\Psi)$ indirectly by altering
the growth rate of $\Delta$ via
\be
\ddot{\Delta}+\mathcal{H}\dot{\Delta}-4\pi{G}\rho{a^2}\mu_0\Delta=0 \, ,
\ee
where the over-dot represents differentiation with respect to the
conformal time, and $\mathcal{H}\equiv\dot{a}/a$ is the conformal
Hubble parameter. Therefore changing $(\Phi+\Psi)$ by varying
$\mu_0$ is much less efficient than tuning the multiplier
$\Sigma_0$. This can be seen in panels ($C_1, C_2$), where we plot
the relative difference of the evolution of $\Delta$ with respect
to GR for different values of $\mu_0$. One can read from the plot
that, for example in the case $z_s=2$, if one fixes $\Sigma_0=1$ and increases $\mu_0$ by 50\%, $\Delta$ is enhanced by 15\% at $z=0$.
From this, it follows from Eq.~(\ref{eq:Sigma}) that $(\Phi+\Psi)$ is also
enhanced by 15\% at $z=0$. Finally $\Map^{1/2}$ should be enhanced
by roughly the same amount according to Eq.~(\ref{eq:Map}), and this
is what we see in panel ($B_2$)~(navy dashed-dotted line). On the other
hand, if $\mu_0$ is fixed to unity and $\Sigma_0$ is enhanced by
50\%, then Eq.~(\ref{eq:Sigma}) clearly shows that $(\Phi+\Psi)$ and
$\Map^{1/2}$ should be enhanced by 50\% as well, as shown in panel
($B_2$) (magenta short-dashed line). Therefore we can conclude that our observables are more sensitive
to $\Sigma_0$ than $\mu_0$ for both cases of the transition redshift, $z_s=1$ and $z_s=2$.

When varying the growth rate controller $\mu_0$, the earlier the redshift at which it is turned on,
the more total change in growth and gravitational potential will be accumulated by the present day, as shown in panels ($C_1,C_2$).
Thus our observables are more sensitive to the same amount of variation in the MG
parameters in the case of $z_s=2$ compared to $z_s=1$ due to this ``accumulation
effect''.

In Fig.~\ref{fig:CMB_imprint}, we see that the CMB angular spectrum will strongly
disfavor a sharp transition in $\Sigma$. This is obvious from
Eq.~(\ref{eq:Sigma}) --- a sharp transition in $\Sigma$ triggers a
sudden change in $(\Phi+\Psi)$, making $(\dot{\Phi}+\dot{\Psi})$
diverge around the transition region, and this is what we see in
panels ($A_1,B_1$). Note that if the transition width $\Delta{z}$ is
small enough, the ISW signal in the CMB spectrum converges because
the details of the transition become irrelevant in the $\Delta{z}
\to 0$ limit; then the ISW signal is determined solely by the change
of $(\Phi+\Psi)$ and the window function at the transition. We find
that the choice $\Delta{z}=0.05$ is narrow enough to approximate
well to this $\Delta{z} \to 0$ limit. We also show the results for a
milder transition, namely, $\Delta{z}=0.5$ in panels ($A_2,B_2$) of
Fig.~\ref{fig:CMB_imprint}. We see that the $\mu_0$ curves are
largely unchanged, implying that the effect of varying $\mu_0$ is
not sensitive to $\Delta{z}$. However, the `bumps' on the $\Sigma_0$
curves become less pronounced for smoother transitions, as
expected. In the following analysis, we will show results corresponding to the
parametrization $\XI$ for both fixed $\Delta{z}=0.05$, and a
floating $\Delta{z}$, which will then be treated as a nuisance
parameter and marginalized over.

\subsection{Pixellation + PCA:  $\mathcal{X}_{\rm II}=\{\mu_i,\Sigma_i,(i=1..4)\}$}\label{sec:PCA}

Even though the parametrization $\mathcal{X}_{\rm I}$ has the
advantage of simplicity,  it is not theoretically
well-motivated. Models of modified gravity commonly introduce a scale into the theory, and correspondingly
$\mu$ and $\eta$ always have some scale dependence.
Moreover, this parametrization is not phenomenologically
efficient to capture a deviation from GR. As shown in \cite{Zhao:2009fn}, the growth observables
are much more sensitive to the scale dependence of the two functions, than to their time dependence.

In order to be more general, one can
pixelize $\mu$ and $\eta$ in the $(k,z)$ plane and treat their values on each grid point as free parameters. These parameters are in general correlated
with each other, and this blurs the interpretation if one attempts to constrain them directly.
Instead, one can construct new variables that are uncorrelated linear combinations of the original parameters and use
them to test GR. This can be achieved by diagonalizing the
covariance matrix of the original pixels and using
the de-correlation matrix to map the original pixels onto the uncorrelated variables.  Such de-correlation, or PCA,
has been used to study constraints on the evolution of the dark energy equation of state
$w(z)$~\cite{PCA1,PCA2,Zhao:2009ti}. Here we
employ a two-dimensional PCA in the $(k,z)$ plane since
$\mu$ and $\eta$ are functions of time and scale.

The model-dependence disappears in the limit of a very fine tessellation of the two functions into pixels. In reality, computing costs limit the number of pixels one can afford to fit. To determine the optimal number of pixels, we performed a Fisher Matrix PCA forecast, analogous to the one in \cite{Zhao:2009fn}, finding that  in order to capture the $(k,z)$ dependence of the best constrained combined eigenmodes one needs at least a $2\times2$ pixels for $\mu$
and $\Sigma$ in the range of $k\in[0,0.2],z\in[0,2]$, as illustrated
in panel A of Fig.~\ref{fig:PCA}.  Note that here we choose to work with the
$\{\mu,\Sigma\}$ parametrization because the ISW and WL constrain $\Sigma$ more directly than $\eta$.
We have checked that our results do not change if we pixelize $\{\mu,\eta\}$ instead. To properly
set the transition width between the neighboring pixels, we start
from a wide transition width ($\Delta{z}=0.5$), and reduce it until
the final results converge. As in the case of parametrization $\XI$,
we found convergent results when $\Delta{z}\lesssim0.05$, therefore
we chose $\Delta{z}=0.05$ for the transition width.

Thus, in model $\mathcal{X}_{\rm II}$, we start by fitting $8$ pixels, $\{\mu_i,\Sigma_i,(i=1..4)\}$, along with the non-MG parameters, to obtain the covariance matrix of all parameters. We then diagonalize the $8 \times 8$ block of the covariance matrix, $C_{(\mu,\Sigma)}$ corresponding to $\mu$ and $\Sigma$:
\be
\label{eq:PCA}
C_{(\mu,\Sigma)}=W\Lambda^{-1}{W}^T \ .
\ee
The rows of the de-correlation matrix $W$ are the principal components~\cite{tegmark-pca}, or eigenmodes, while
the diagonal elements of $\Lambda$, i.e. the eigenvalues, are the inverses of the variances on the uncorrelated linear combinations of the original pixels. Namely, we use $W$ to
rotate the original parameters, denoted by the vector $\mathbf{p}$, into 
new uncorrelated parameters $\mathbf{q}$ defined as
\be\label{eq:q}
q_i=-1+\sum_jW_{ij}p_j/\sum_jW_{ij} \, .
\ee
In GR one has $\mathbf{q}=\mathbf{0}$, since $\mathbf{p}=\mathbf{1}$, therefore we
can test GR by performing a null test on $\mathbf{q}$. By construction, the
eigenvectors are orthogonal and the $q$'s have uncorrelated errors given by the inverses of the eigenvalues.

\section{Results} \label{sec:results}

\begin{table*}
\caption{The mean values of $\mu_0,~\eta_0$ and $\Sigma_0$ with 68\%
and 95\% C.L. error bars for different models and for different data
combinations. Note that `$\ell_A,R$',`CMB', `ISW' and `WL' are
short-hands for WMAP5 shift parameters, full WMAP5 data, ISW data and CFHTLS data explained in the text, respectively.} \vspace{0.5cm}
\centering
\begin{tabular}{c|c|c|c|c|c|c|c}
\hline\hline
\multicolumn{2}{c|}{} & \multicolumn{3}{c|}{$z_s=1$}& \multicolumn{3}{c}{$z_s=2$}   \\
\hline
\multicolumn{2}{c|}{} & $\mu_0$ & $\eta_0$ & $\Sigma_0$  & $\mu_0$ & $\eta_0$ &$\Sigma_0$   \\
\hline
                   & CMB &$1.0^{+0.11+0.40}_{-0.13-0.34}$ & $1.1^{+0.51+1.0}_{-0.48-0.74}$ & $1.0\pm0.03\pm0.06$     & $1.1^{+0.16+0.37}_{-0.17-0.31}$&  $0.96^{+0.11+0.62}_{-0.18-0.47}$
                   &$1.0\pm0.025\pm0.05$\\
                   & CMB$+$ISW &$0.97^{+0.09+0.37}_{-0.13-0.29}$ & $1.2\pm0.50^{+0.94}_{-0.77}$ & $1.1\pm0.028\pm0.055$     & $1.0^{+0.15+0.35}_{-0.16-0.28}$&  $0.98^{+0.10+0.55}_{-0.17-0.45}$ &$1.0\pm0.024\pm0.05$ \\
$\Delta{z}$ fixed  & WL$+\ell_A,R$ &$0.63^{+0.65+1.36}_{-0.45-0.57}$ & $1.7^{+3.2+6.2}_{-1.6-2.3}$ & $0.86\pm0.39\pm0.74$     & $0.58^{+0.92+1.17}_{-0.38-0.55}$&  $2.1^{+3.0+5.7}_{-1.5-2.2}$ &$0.89\pm0.19\pm0.33$ \\
                   &WL$+$CMB &$0.95\pm0.24^{+0.54}_{-0.34}$ &  $1.2\pm0.60^{+1.1}_{-0.91}$  & $1.0\pm0.033^{+0.07}_{-0.06}$ &$0.87\pm{0.15}^{+0.33}_{-0.25}$ & $1.4\pm{0.44}^{+0.91}_{-0.76}$ &$1.0\pm0.03\pm0.05$  \\
                   & WL$+$CMB$+$ISW & $0.90\pm0.21^{+0.42}_{-0.29}$ & $1.3\pm{0.56}^{+0.98}_{-0.84}$ & $1.0\pm0.027\pm0.05$ & $0.84\pm0.13^{+0.27}_{-0.21}$ &$1.4\pm0.39^{+0.81}_{-0.69}$ &$1.0\pm0.025\pm0.05$   \\

\hline

  $\Delta{z}$ float & WL$+$CMB$+$ISW & $1.1^{+0.62+0.80}_{-0.34-0.45}$ & $0.98^{+0.73+1.2}_{-1.0-1.4}$ & $0.94^{+0.08+0.12}_{-0.14-0.32}$  & $0.87\pm0.12^{+0.24}_{-0.19}$  &$1.3\pm0.35^{+0.65}_{-0.60}$  &$1.0\pm0.03\pm0.06$  \\

\hline\hline
\end{tabular}
\label{tab:const}
\end{table*}

\begin{figure} [htp]
\begin{center}
\includegraphics[scale=0.7]{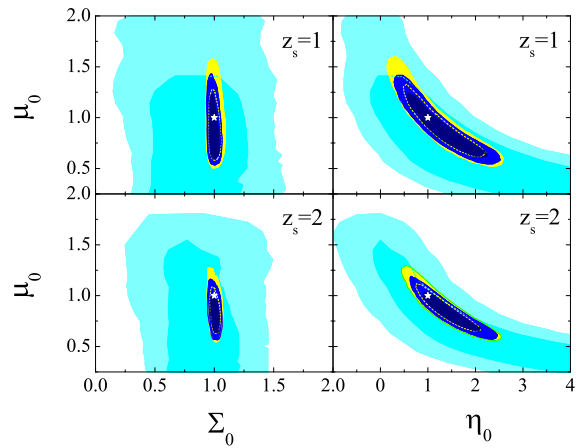}
\caption{68\% and 95\% C.L. contour plots for $\{\mu_0,~\eta_0\}$ and
$\{\mu_0,~\Sigma_0\}$ for two different threshold redshifts: $z_s=1$
(upper panels) and $z_s=2$ (lower panels). In both cases the
transition width is fixed to $\Delta{z}=0.05$. From outside in, the
shaded regions in cyan, yellow and blue illustrate the contours
derived from the data of CFHTLS+CMB shift parameters, CFHTLS+WMAP5
and CFHTLS+WMAP5+ISW, respectively. For the contours shaded in the
same color, the light and dark regions show the 68\% and 95\% C.L.
contour respectively. In all cases, the SNe data are combined, and
the priors of cosmic age, BBN and HST are applied. The star
denotes the GR values.} \label{fig:cont1}
\end{center}
\end{figure}

\begin{figure} [htp]
\begin{center}
\includegraphics[scale=0.75]{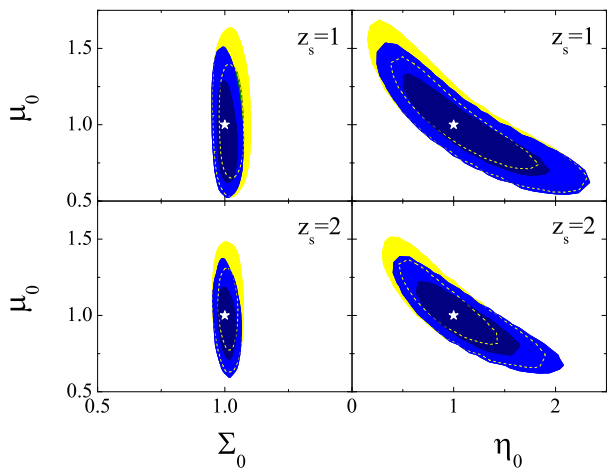}
\caption{ 68\% and 95\% C.L. contour plots for $\{\mu_0,~\eta_0\}$ and
$\{\mu_0,~\Sigma_0\}$ for two different threshold redshifts: $z_s=1$
(upper panels) and $z_s=2$ (lower panels). In both cases the
transition width is fixed to $\Delta{z}=0.05$. From outside in, the
shaded regions in yellow and blue illustrate the contours
derived from the data of WMAP5 and WMAP5+ISW, respectively.
For the contours shaded in the same color, the light and dark regions
show the 68\% and 95\% C.L.
contour respectively. In all cases, the SNe data are combined, and
the priors of cosmic age, BBN and HST are applied. The star
denotes the GR values.}\label{fig:cont3}
\end{center}
\end{figure}

\begin{figure} [htp]
\begin{center}
\includegraphics[scale=0.7]{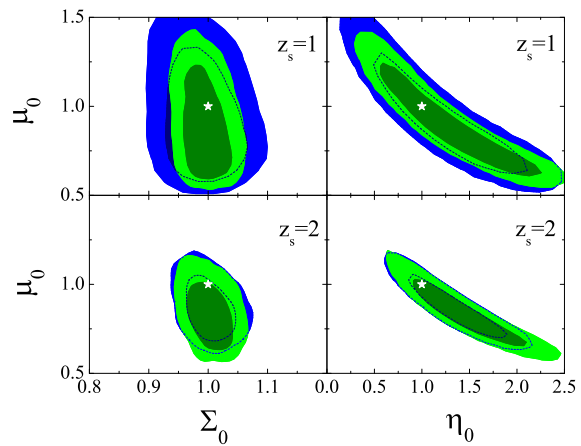}
\caption{68\% (dark shaded) and 95\% C.L. (light shaded) contour
plots for $\{\mu_0,\Sigma_0\}$ and $\{\mu_0,\eta_0\}$ for two
different threshold redshifts: $z_s=1$ (upper panels) and $z_s=2$
(lower panels). All the constraints are from the combined data of
ISW, WMAP5 and CFHTLS. To obtain the front, green contours, the
transition width is fixed to $\Delta{z}=0.05$, while the blue
contours on the back layers show the case of a floating $\Delta{z}$,
which is marginalized over. The dashed curves show the covered
contour edges. The star illustrates the GR values.}
\label{fig:cont2}
\end{center}
\end{figure}

\begin{figure} [htp]
\begin{center}
\includegraphics[scale=0.6]{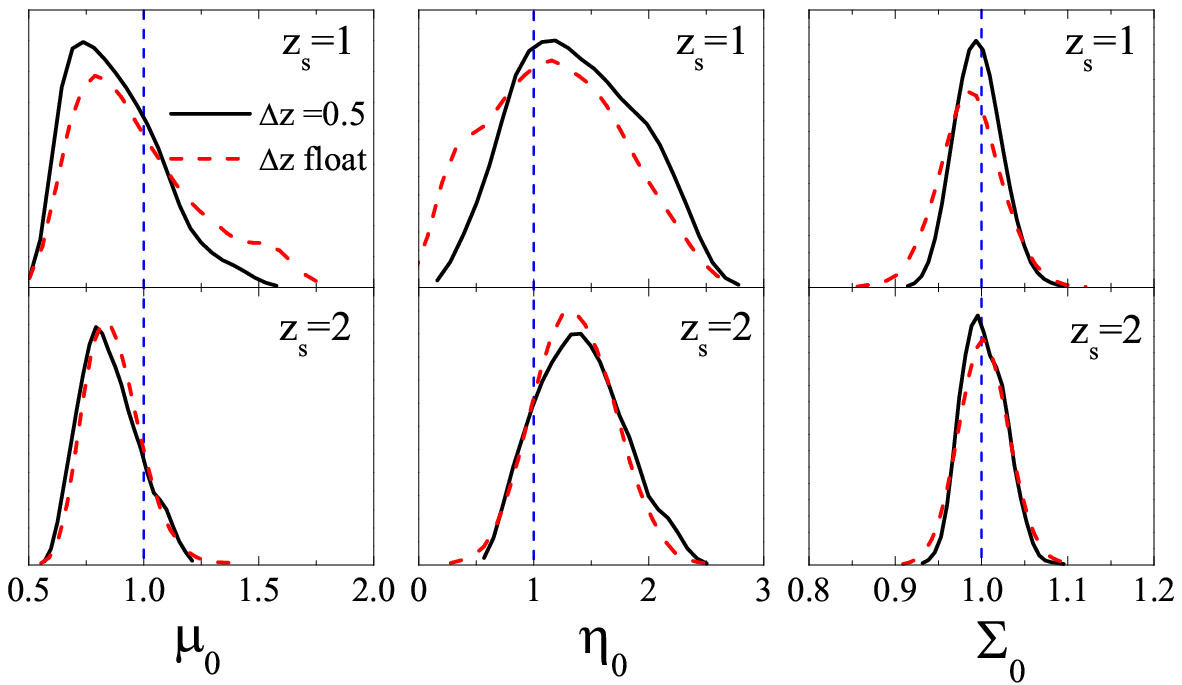}
\caption{1-D posterior distributions of $\mu_0, \eta_0$, and
$\Sigma_0$ derived from the joint analysis of ISW, WMAP5 and CFHTLS
data. The black solid lines show the cases of sharp transition,
i.e., $\Delta{z}=0.05$, while the red dashed lines illustrated the
cases where $\Delta{z}$ varies and is marginalized over. The upper
and lower panels are for the $z_s=1$ and $z_s=2$ cases,
respectively. The vertical dashed lines illustrate
$\mu_0=\eta_0=\Sigma_0=1$ to guide eyes.} \label{fig:1D}
\end{center}
\end{figure}

\begin{figure*} [htp]
\begin{center}
\includegraphics[scale=0.8]{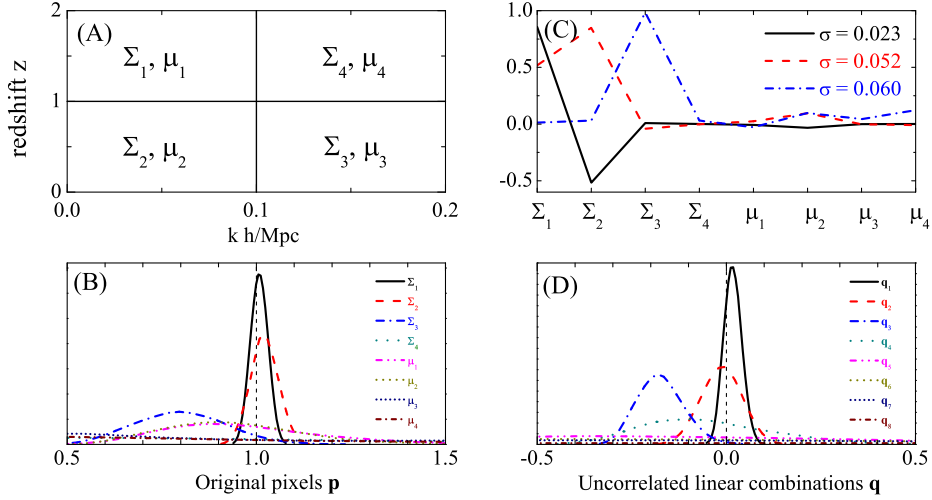}
\caption{Panel (A): the $\Sigma,\mu$ pixellation used in this work;
Panels (B) and (D): the 1-D posterior distributions of $\mathbf{p}$
and $\mathbf{q}$, which denote the original pixels of $\Sigma$ and
$\mu$, and the uncorrelated linear combinations of the original
pixels, respectively. The likelihood distributions are normalized so
that the area of each distribution is unity; Panel (C): the first
three eigenfunctions, i.e. values of $\mathbf{W}$ relating
$\mathbf{p}$ and $\mathbf{q}$ via Eq.~(\ref{eq:q}).
These are derived from the joint analysis of ISW, WMAP5 and CFHTLS data.
} \label{fig:PCA}
\end{center}
\end{figure*}

Given the set of cosmological parameters ${\bf P}$ in
Eq.~(\ref{eq:paratriz}), we calculate the observables, which include
the CMB temperature and
polarization spectra, the CMB/galaxy cross-correlation (which we often refer to as ISW), the luminosity
distance for SNe, and the WL aperture-mass dispersion $\Map$, using {\tt MGCAMB}. We then fit the available CMB, ISW, SNe and WL to observations using a modified version of the Markov
Chain Monte Carlo (MCMC) package {\tt
CosmoMC}~\footnote{\url{http://cosmologist.info/cosmomc/}}~\cite{CosmoMC},
based on Bayesian statistics. Our main results are summarized in
Figs.~\ref{fig:cont1}-\ref{fig:PCA2} and Tables~\ref{tab:const}
-\ref{tab:PCA2}.

\subsection{Parametrization $\XI$}

Let us start with the single-transition parametrization $\mathcal{X}_{\rm
I}$ of the MG parameters. In Fig.~\ref{fig:cont1}, we show the 68\%
and 95\% C.L. contours of $\{\mu_0,~\eta_0\}$ and
$\{\mu_0,~\Sigma_0\}$ for the cases of $z_s=1$ (upper panels) and
$z_s=2$ (lower panels) for different data combinations. Here the
transition width is fixed to $\Delta{z}=0.05$. We show contours
derived from the CFHTLS data combined with the CMB shift parameters
$\ell_A$ and $R$ given in \cite{Komatsu:2008hk}, CFHTLS plus full
WMAP5 and CFHTLS+WMAP5+ISW. In Fig.~\ref{fig:cont3}, we also
show contours derived from full WMAP5 and WMAP5 + ISW.
All cases include the SNe data, and the cosmic age, BBN and HST priors.

One thing that can be noticed from Figs.~\ref{fig:cont1} and \ref{fig:cont3} is that the $z_s=2$
models, shown in the lower panels, are in general better constrained than
the $z_s=1$ ones, shown in the upper panels. This is due to the
`accumulation effect' explained in Sec.~\ref{sec:XI}. 

In Fig. \ref{fig:cont1}, the largest cyan contours show that the CFHTLS WL data combined with CMB
shift parameters are able to constrain $\Sigma_0$ at $\sim$
20\% level, but only weakly constrain $\mu_0$ and $\eta_0$, since WL observables are directly sensitive to the variation in $\Sigma_0$ as explained in Sec.~\ref{sec:XI}. For the same reason, one
sees little degeneracy in the $\{\Sigma_0,~\mu_0\}$ plane. However,
in the $\{\mu_0,~\eta_0\}$ plane, the contours show a banana shape,
which indicates that $\mu_0$ strongly anti-correlates with $\eta_0$. This is
understandable --- one can increase $\mu_0$ to enhance growth and thus $\Psi$,
but then $\eta_0$ can be lowered to decrease $\Phi$, leaving
$\Phi+\Psi$ unchanged. Also, as discussed in Sec.~\ref{sec:XI}, the sensitivity of the observables to $\mu_0$ is comparable
to their sensitivity to $\eta_0$, although the former is slightly larger than the latter.
This is the reason why the degeneracy is visible. From Table I, we
see that CFHTLS combined with CMB shift parameters favor slightly
lower values of $\mu_0$ and $\Sigma_0$ than unity, but GR is still
within the $1\sigma$ level.

The constraints become much tighter if one includes
the full WMAP5 data, as shown in the yellow contours in Figs.~\ref{fig:cont1} and \ref{fig:cont3}.
This is mainly because the WMAP5 ISW-ISW ACFs strongly penalize abrupt
changes in $\Sigma_0$, as we show in Fig.~\ref{fig:CMB_imprint}. The
constraints get even tighter when the ISW-gal CCFs data are added, as
illustrated in the innermost blue contours. 
From Fig.~\ref{fig:cont1} and Table I we see that for the case of
$z_s=1$ GR is fully consistent with the combined
data. For the $z_s=2$ case, GR is also consistent, but is on the $1\sigma$ edge, indicating that a model with
a lower $\mu_0$ would be slightly favored when WL data are included. GR is always closer to the best-fit model when WL data are not used, as we can see in Fig.~ \ref{fig:cont3}.

If the transition width $\Delta{z}$ is allowed to vary in the range
of $[0.05,0.5]$ one could expect a dilution of the
constraints. We observe the result in Figs.~\ref{fig:cont2} and
\ref{fig:1D}, which show the contours and 1-D posterior
distributions for MG parameters for the cases of sharp transitions
and floating transitions for all the data combined. As we can see,
marginalizing over a floating $\Delta{z}$ degrades the constraints
on $\mu_0,\eta_0$ and $\Sigma_0$ by roughly 150\%, 50\% and 300\%,
respectively for the $z_s=1$ case, but there is little degradation
for the $z_s=2$ case. Again, we see that GR is a perfect fit in the
$z_s=1$ case, while a model with a lower $\mu_0$ is slightly favored
in the $z_s=2$ case when all data are combined.

This can be understood as follows. As shown in panels $(A_1,B_1)$ in
Fig.~\ref{fig:CMB_imprint}, for the $z_s=1$ case, a sharp transition
in $\Sigma$ produces a huge bump on CMB TT spectrum at
$\ell\lesssim70$, where the cosmic variance dominates the error
budget. If the transition is mild, the bump structure becomes less
pronounced as shown in panel $A_2$, thus there is less tension with
the CMB data, which in turn loosens the constraints on the MG
parameters. However, for the $z_s=2$ case, the bump appears at
$\ell\lesssim150$ on the CMB spectrum, where WMAP5 has precise
measurements. It is true that relaxing $\Delta{z}$ reduces the bump
feature somewhat as illustrated in panel $B_2$; however the
constraints on the MG parameters cannot be diluted to a large
extent due to the high quality CMB data at
$70\lesssim\ell\lesssim150$.

Note that the constraint on $\Sigma$ from the
ISW effect is much tighter than the current WL constraints, due to the sensitivity of the ISW to the gravitational transition. This means that the ISW data provide  valuable information on the time evolution
of modified gravity parameters.

\subsection{Parametrization $\XII$}

\begin{figure} [htp]
\begin{center}
\includegraphics[scale=0.75]{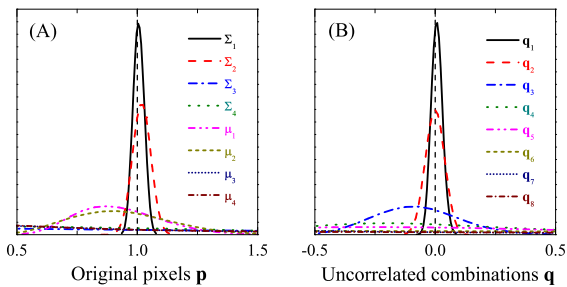}
\caption{
Panels (A) and (B): The 1-D posterior distributions of $\mathbf{p}$
and $\mathbf{q}$, which denote the original pixels of $\Sigma$ and
$\mu$, and the uncorrelated linear combinations of the original
pixels, respectively. The likelihood distributions are normalized so
that the area of each distribution is unity. 
These are derived from the joint analysis of ISW and WMAP5 data, without the CFHTLS WL data..}\label{fig:PCA2}
\end{center}
\end{figure}

We present results for the second parametrization $\mathcal{X}_{\rm II}$ with and without the inclusion of WL data, as summarized in Figs.~\ref{fig:PCA},~\ref{fig:PCA2} and Tables~\ref{tab:PCA},~\ref{tab:PCA2} respectively.
Starting from the analysis based on the full data set, in
panel (A) of Fig.~\ref{fig:PCA}, we show our $\Sigma$ and $\mu$
pixellation, and in panel (B), we show the 1-D posterior
distributions of the eight $\Sigma$ and $\mu$ pixels. As we found
for the parametrization $\mathcal{X}_{\rm I}$, the $\Sigma$ pixels
are in general better constrained than the $\mu$ pixels.
We find that the constraints on all the pixels are
consistent with the GR prediction except for that of $\Sigma_3$,
which is $\Sigma_3=0.80\pm0.12\pm0.22$ (mean value with 68\% and
95\% C.L. errors). This means that $\Sigma_3$ deviates from the GR
prediction at an almost $2 \sigma$ level. However, the correlation
among all of the eight pixels blurs the na\"{i}ve interpretation of
the seemingly $2 \sigma$ signal. Thus, we follow the PCA
prescription explained in Sec.~\ref{sec:PCA} and obtain the linearly
uncorrelated parameters, $q$'s, using Eq.~(\ref{eq:q}).

The three best constrained eigenmodes are shown in panel $(C)$ of
Fig.~\ref{fig:PCA} and are fairly well-localized. One can clearly see that they primarily depend on the $\Sigma$ pixels, which are expected to be better measured by the ISW and WL. In particular, the eigenmode corresponding to $q_i~(i\leq3)$ received the largest contribution from $\Sigma_i~(i\leq3)$. From the eigenmodes we can deduce the following relations between the $q's$ and the original pixels:
\ba\label{eq:q123}
&&q_1\approx-1+\frac{0.85\Sigma_1-0.52\Sigma_2}{0.85-0.52}=0.0\pm0.02\pm0.04\nonumber\\
&&q_2\approx-1+\frac{0.52\Sigma_1+0.85\Sigma_2}{0.52+0.85}=0.0\pm0.05\pm0.10\nonumber\\
&&q_3\approx-1+\Sigma_3=-0.17\pm0.06^{+0.13}_{-0.11} \, .
\ea
Namely, $\Sigma_1$ is strongly degenerate with $\Sigma_2$, while $\Sigma_3$ is largely independent of
$\Sigma_1$ or $\Sigma_2$;

\begin{table}
\caption{Mean values, and 68\% and 95\% C.L. constraints of the
original pixels (left panel) and the uncorrelated linear
combinations of the pixels (right panel). All data sets are used.} \vspace{0.5cm}

\begin{tabular}{cc|cc}
\hline \hline
$\Sigma_1$               &$1.0\pm0.02\pm0.04$       &$q_1$      &$0.0\pm0.02\pm0.04$    \\
$\Sigma_2$               &$1.0\pm0.04\pm0.07$       &$q_2$      &$0.0\pm0.05\pm0.10$   \\
$\Sigma_3$               &$0.80\pm0.12\pm0.22$           &$q_3$      &$-0.17\pm0.06^{+0.13}_{-0.11}$   \\
$\Sigma_4$               &$0.83^{+0.63+1.4}_{-0.60-0.83}$           &$q_4$      &$-0.05\pm0.17^{+0.37}_{-0.28}$   \\

$\mu_1$               &$0.96\pm0.20^{+0.46}_{-0.32}$      &$q_5$      &$-0.10\pm0.52^{+1.1}_{-0.81}$   \\
$\mu_2$               &$0.94\pm0.18^{+0.40}_{-0.29}$       &$q_6$      &$-0.17\pm0.79^{+1.7}_{-1.2}$   \\
$\mu_3$               &$0.94^{+0.64+1.3}_{-0.52-0.70}$           &$q_7$      &$-0.02^{+1.1+2.1}_{-1.0-2.0}$   \\
$\mu_4$               &$0.86^{+0.69+1.6}_{-0.62-0.81}$           &$q_8$      &$-0.25\pm3.2^{+6.0}_{-5.2}$   \\

\hline  \hline
\end{tabular}
\label{tab:PCA}
\end{table}

\begin{table*}
\caption{The relative improvement on $\chi^2$ with respect to the
$\Lambda$CDM model for different models. The $\Delta\chi^2$ is shown
for different data separately. The mean values with 68\% and 95\% C.L.
error bars of constraints on $\Omega_{\rm m}$ and $\sigma_8$ are
also shown.} \vspace{0.5cm} \centering
\begin{tabular}{c|ccccc|cc}
\hline\hline

&$\Delta\chi^2_{\rm WL}$&$\Delta\chi^2_{\rm CMB}$&$\Delta\chi^2_{\rm ISW}$&$\Delta\chi^2_{\rm SN}$&$\Delta\chi^2_{\rm
ToT}$&$\Omega_{\rm m}$&$\sigma_8$\\
\hline
$\Lambda$CDM&0&0&0&0&0&$0.244\pm0.004$&$0.765\pm0.006$\\
$\mathcal{X}=\XI,~z_s=1,~\Delta{z}=0.05$&$-0.65$&$+0.11$&$-0.10$&$-0.17$&$-0.81$&$0.252\pm0.008$&$0.74\pm0.02$\\
$\mathcal{X}=\XI,~z_s=1,~\Delta{z}$ float&$-0.63$&$+0.01$&$-0.18$&$-0.50$&$-1.3$&$0.251\pm0.008$&$0.74\pm0.02$\\
$\mathcal{X}=\XI,~z_s=2,~\Delta{z}=0.05$&$-0.73$&$+2.0$&$-0.45$&$-2.4$&$-1.6$&$0.247\pm0.006$&$0.76\pm0.02$\\
$\mathcal{X}=\XI,~z_s=2,~\Delta{z}$ float&$-1.4$&$+1.8$&$-0.44$&$-2.6$&$-2.6$&$0.255\pm0.019$&$0.79\pm0.06$\\
$\mathcal{X}=\XII$, all pixels float&$-2.3$&$+4.1$&$+1.4$&$-12.3$&$-9.1$&$0.30\pm0.024$&$0.80\pm0.069$\\
$\mathcal{X}=\XII$, only $\Sigma_3$ float&$-2.2$&$+5.3$&$+0.48$&$-12.2$&$-8.6$&$0.30\pm0.022$&$0.82\pm0.021$\\

\hline\hline
\end{tabular}
\label{tab:chi2}
\end{table*}

\begin{figure} [htp]
\begin{center}
\includegraphics[scale=1.]{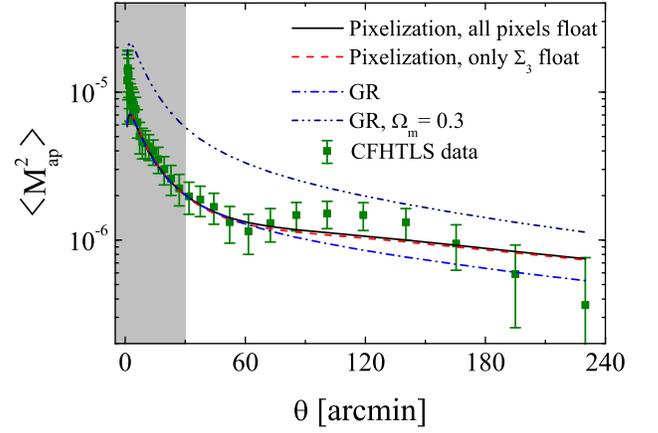}
\caption{Best-fit aperture mass power spectra $\langle{M_{\rm
ap}^2}\rangle$ for different MG parametrizations shown in different
colors and line styles. Black solid: parametrization
$\mathcal{X}_{\rm II}$ with all the pixels varying; Red dashed:
parametrization $\mathcal{X}_{\rm II}$ with only $\Sigma_3$ varying;
Blue dashed-dotted: GR; Navy dashed-dotted-dotted:  GR with a
fixed $\Omega_m=0.3$. The data points with error bars show the
CFHTLS data; the shaded region is excluded from our analysis.} \label{fig:CFHTLS_bf}
\end{center}
\end{figure}

\begin{figure} [htp]
\begin{center}
\includegraphics[scale=0.6]{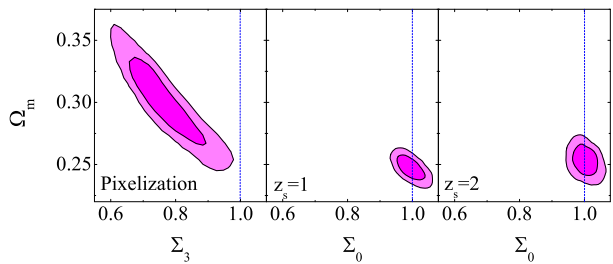}
\caption{The 68 and 95\% C.L. contour plots of $\{\om, \Sigma_3\}$
(left panel) for parametrization $\XII$, and $\{\om, \Sigma_0\}$
(middle and right panels) for parametrization $\XI$. See text for
details.} \label{fig:om_Sigma}
\end{center}
\end{figure}

One can understand this by realizing that the current ISW data can put stronger constraints
on the MG parameters than the current WL data, due
to the sensitivity of the ISW to any modification of growth at
$z\lesssim2$. Future WL surveys (including the upcoming final results from the full CFHTLS survey) will feature higher signal-to-noise and will provide tighter limits on MG parameters~\cite{Zhao:2009fn}; however, at present, the ISW component dominates in our data combination, making the linear
combinations of the $\Sigma$ pixels on large scales
($k\lesssim0.1$h/Mpc), $\Sigma_1$ and $\Sigma_2$, best measured.

The constraints on the $q$'s are summarized in the right panel of
Table~\ref{tab:PCA}, and the 1-D posterior distributions are shown
in panel (D) of Fig.~\ref{fig:PCA}. We find that all the $q$'s are
consistent with zero as predicted by GR, except for $q_3$,
which deviates from the GR prediction at more than 95\% confidence
level. This means that our measurement of $\Sigma$ and/or $\mu$ deviate
from unity at the level of at least 95\% C.L. at some point in $(k,z)$.

To spot the source of this signal, we fix all the pixels to unity
except for $\Sigma_3$, which we allow to vary. Interestingly, we
find that the goodness-of-fit of this one-pixel model is almost
identical to that of the 8-pixel model. To be explicit, we list the
$\chi^2$ for each dataset separately for both models, and also show
the constraints on $\Omega_{\rm m}$ and $\sigma_8$ for these models
in Table~\ref{tab:chi2}. For comparison, we also do the same for the
models parametrized using $\XI$. Comparing to $\Lambda$CDM, we find
that allowing $\Sigma_3$ to vary can reduce the WL $\chi^2$ by $2.2$, while also reducing the SNe $\chi^2$ by roughly $12$. This $\Sigma_3$CDM model has a best-fit $\Omega_{\rm m}$ of $0.3$, which
is much larger than $0.24$ for $\Lambda$CDM. The
allowance for a high $\Omega_{\rm m}$ is the reason for the
significant SNe data preference for this one-pixel model, since
$\Omega_{\rm m}=0.3$ is the best-fit value for the SNe sample we
use~\cite{Kessler:2009ys}. Note that in $\Lambda$CDM, $\Omega_{\rm
m}=0.3$ is strongly disfavored by WL data, as we show in
Fig.~\ref{fig:CFHTLS_bf} (navy dashed-dotted-dotted line), since
there increasing $\Omega_{\rm m}$ shifts the best-fitted $\Map$
(blue dashed-dotted) on all scales, which is in serious disagreement
with the data, especially on scales $\theta<60$ arc min.

The cause of the apparent $2\sigma$ hint of departure from GR can be easily identified. There is a clear ``bump''
feature in the CFHTLS data (e.g. Fig.~{\ref{fig:CFHTLS_bf}) at $\theta
  \simeq 120 $ arc min, which can be attributed to a systematic
  effect~\cite{wl_private}:
according to the CFHTLS team, this is a known issue, due to residual field-to-field variations in shear estimation on the scale of the camera field-of-view.
 As an informative exercise, we study how we could improve the fit assuming a cosmological source for the feature. One could shift the  curve at $\theta\geq60$ arc min to follow the
``bump'' more closely. Such a scale-dependent tweak of $\Map$ can not be
realized by tuning the MG parameters in the parametrization $\XI$.
However, in the parametrization $\XII$, one can achieve this by firstly
increasing $\Omega_{\rm m}$, then lowering the growth rate
on small scales ($\theta<60$ arc min), which can be effectively done
by lowering $\Sigma_3$. The resultant fit is shown in Fig.~\ref{fig:CFHTLS_bf} as a red-dashed
line, which is almost identical to the best-fit 8-pixel model.

To see this point more clearly, in Fig.~\ref{fig:om_Sigma} we show the contour plots between
$\Omega_{\rm m}$ and $\Sigma_3$, and $\Omega_{\rm m}$ and $\Sigma_0$ of $\XI$ for
the combined data. As we can see, in
the one-pixel $(\Sigma_3)$ model can one obtain a high $\Omega_{\rm m}$ as favored by
the SNe data, while in parametrization $\XI$, $\Sigma_0$ is
tightly constrained, so that a high $\Omega_{\rm m}$ is definitely
not allowed. Notice that the CMB and ISW data disfavor a high
$\Omega_{\rm m}$, as we see in Table~\ref{tab:chi2}, but the
preference from the WL and SNe data outweighs this penalty, making a
lower $\Sigma_3$ and higher $\Omega_{\rm m}$ strongly favored by
the combined data. Also, the one-pixel model would be strongly favored by the
data from the model-selection point of view, since one can reduce the
total $\chi^2$ by $8.6$ by introducing one more parameter $(\Sigma_3)$
over $\Lambda$CDM, even though this model was constructed \emph{a posteriori}.

As stated above, it is likely that the ``bump'', which is responsible
for this $2\sigma$ deviation, is due to a systematic
effect~\cite{wl_private}, which according to the CFHTLS team
  is due to residual field-to-field variations in shear estimation
  on the scale of the camera field-of-view; this explains the scale
  of the bump. On these grounds, we stress again that it is premature
  to make any statements about the validity of $\Lambda$CDM based on this
  feature, even though 
  technically we cannot rule out the new physics at this point.

In order to be conservative about this issue, we also perform the $\mathcal{X}_{\rm II}$ analysis without including the WL data from CFHTLS. The results are summarized in Fig.~\ref{fig:PCA2} and Table~\ref{tab:PCA2}. The relation between the uncorrelated parameters 
$q$'s and the original pixels, and the 68$\%$ and 95$\%$ C.L. constraints on the $q$'s are now given by 
\ba\label{eq:q123B}
&&q_1\approx-1+\frac{0.90\Sigma_1-0.44\Sigma_2}{0.90-0.44}=0.0\pm0.02\pm0.04\nonumber\\
&&q_2\approx-1+\frac{0.44\Sigma_1+0.90\Sigma_2}{0.44+0.90}=0.0\pm0.04\pm0.07\nonumber\\
&&q_3\approx-1+\frac{0.52\mu_1-0.85\mu_2}{0.52-0.85}=-0.07\pm0.17^{+0.36}_{-0.31}
\, . \ea
As expected, the first two best constrained modes are almost unchanged even if we remove the CFHTLS data, 
confirming that these modes are mostly constrained by the ISW effect. On the other hand, the bound
on $\Sigma_3$ becomes very weak, demonstrating that the CFHTLS data are 
responsible for the constraint on this parameter. In the current case we find that all the eight
uncorrelated parameters are consistent with GR with 95 $\%$ C.L. }

\begin{table}
\caption{Mean values, and 68\% and 95\% C.L. constraints of the
original pixels (left panel) and the uncorrelated linear
combinations of the pixels (right panel). Results without the CFHTLS WL data.} \vspace{0.5cm}

\begin{tabular}{cc|cc}
\hline \hline
$\Sigma_1$               &$1.0\pm0.02\pm0.04$       &$q_1$      &$0.0\pm0.02\pm0.04$    \\
$\Sigma_2$               &$1.0\pm0.04\pm0.07$       &$q_2$      &$0.0\pm0.04\pm0.07$   \\
$\Sigma_3$               &$1.2^{+1.0+2.1}_{-0.90-1.2}$           &$q_3$      &$-0.07\pm0.17^{+0.36}_{-0.31}$   \\
$\Sigma_4$               &$0.84^{+0.62+1.3}_{-0.60-0.83}$           &$q_4$      &$-0.06\pm0.43^{+0.90}_{-0.69}$   \\

$\mu_1$               &$0.93\pm0.17^{+0.38}_{-0.28}$      &$q_5$      &$-0.12\pm0.62^{+1.2}_{-1.0}$   \\
$\mu_2$               &$0.93\pm0.20^{+0.44}_{-0.34}$       &$q_6$      &$0.30^{+1.3+2.7}_{-1.3-1.9}$   \\
$\mu_3$               &$0.95^{+0.85+1.8}_{-0.71-0.91}$           &$q_7$      &$-0.17^{+1.5+3.4}_{-1.1-1.8}$   \\
$\mu_4$               &$0.89^{+0.76+1.7}_{-0.66-0.84}$           &$q_8$      &$-0.31^{+1.9+4.1}_{-1.7-4.0}$   \\

\hline  \hline
\end{tabular}
\label{tab:PCA2}
\end{table}

\section{Conclusion and Discussion}  \label{sec:concl}

We have tested GR with current cosmological data, using a framework in which the departures from GR are encoded as modifications of the
 anisotropy and Poisson equations; these equations specify, respectively, how the metric perturbations relate to each other, and how they are sourced by perturbations in the energy-momentum tensor of matter. The modifications were parametrized with
two functions $\{\eta, \mu\}$ (or alternatively $\{\Sigma, \mu\}$) that reduce to unity in GR. We
have then explored the constraints on these functions in two ways. First, we have allowed them to evolve from unity at high redshifts to a different value today in a scale-independent way. Second, we
have \emph{pixelized them in both scale and redshift} and performed a  Principal Component Analysis (PCA), following the ideas of \cite{Pogosian:2010tj,Zhao:2009fn} --- a first general study of this kind.
Specifically, we have used a $2 \times 2$ pixellation for each function, thus having
$8$ modified gravity parameters. In order to remove the
covariance between the bins, and to analyze which modes are best
constrained, we have then performed a 2D PCA of the results,
obtaining constraints on the  $8$ derived de-correlated
parameters.

We have used currently available data constraining both the
background expansion history and the evolution of scalar perturbations in the Universe.
In particular, we have used a combined measurement of the ISW
effect through correlation of galaxies with CMB, the latest available supernovae Type Ia
data including those from the SDSS, the CMB temperature and
polarization spectra from WMAP5, and weak lensing data from the
CFHTLS shear catalog. We have kept the analysis conservative by excluding small-scale data in the strongly non-linear regime, and we have checked and excluded possible
tensions between the data sets by analyzing them individually
before combining them.

Throughout the paper, we have assumed a flat $\Lambda$CDM background and
tried to constrain deviations from GR in the evolution of matter and metric
perturbations. In the simplest case, where the MG functions $\{\mu,
\Sigma\}$ were allowed a single transition in redshift, we have found no
evidence for a departure from GR, in agreement with other works. We find that the ISW effect, probed through the CMB auto-correlation and the cross-correlation with galaxy maps, currently gives the strongest
constraint on $\Sigma$ because it is sensitive to the change of the lensing potential,
$\Phi + \Psi$, at the transition.

In the pixellated case, we have found that one of the PCA eigenmodes
shows a $2 \sigma$ deviation from GR. However, this anomalous mode is due to the
``bump'' feature in the CFHTLS lensing data, which is most likely due to a systematic effect~\cite{wl_private},  combined with a preference
for higher $\Omega_m$ by the SNe data.
A separate analysis which does not include WL data shows indeed good agreement with GR.
A better understanding of systematic effects in both WL and SNe data sets needs to be achieved before any such discrepancy is viewed as a deviation from GR.

Even though this is most likely due to a known systematic effect, we emphasize that we would
not have found this deviation if
$\Sigma$ were taken to be scale-independent. In such case, the change
in $\Sigma$ would be significantly constrained by the ISW effect.
The PCA analysis using two bins in $k$ for $\Sigma$ could successfully isolate the strong
constraint from the ISW effect and pick up a feature in WL. This
demonstrates that the same data can lead to a higher level of
detection of deviations from an expected model if more flexibility is allowed in
the parametrization. Thus, when fitting $\mu(k,z)$ and $\eta(k,z)$ to
data, it is important that their parametrization allows for some scale-dependence. Otherwise, one might
risk missing a systematic effect or a real departure from GR, and thus would not
be exploiting the true discovery potential of the data.

Finally, we comment on other recent studies that reported constraints on deviations from GR using
current cosmological observations. In \cite{Bean:2010zq}, the COSMOS
weak lensing tomography data \cite{Massey:2007gh} were used together
with SNe, CMB, BAO and the ISW-galaxy cross correlation.
\cite{Daniel:2010ky} found that the constraints from CMB+SNe+CFHTLS without COSMOS were indistinguishable from those including COSMOS and they did not find any deviations from GR. They argued that the sky coverage of CFHTLS is more important than the redshift depth of COSMOS. Also it should be noted that weak lensing measurements in COSMOS are made on strongly non-linear scales and there is an ambiguity in modeling the non-linear power spectrum.

In \cite{Daniel:2010ky}, a similar set of data to that
described here was used to constrain two functions that are
combinations of $\mu$ and $\eta$. The differences between our study and that of \cite{Daniel:2010ky}
include: (1) we used the ISW cross-correlation data; (2) we excluded small-scale modes in the CFHTLS data to avoid the strongly non-linear regime; (3) we simultaneously constrained two functions
$\mu$ and $\eta$ while \cite{Daniel:2010ky} varied only one of the
parameters when they use 3 bins in $z$; (4) scale dependence was
allowed in our paper; and (5) we used the Fisher matrix based PCA
approach to make a decision on how many pixels to use. Their results
are qualitatively consistent with the result of our first parametrization.

Further improvements of this technique will be possible with a new
generation of LSS data (e.g. DES, Pan-STARRS,
LSST, Euclid), which will dramatically increase the number of modes
with sufficient signal-to-noise. Finally, peculiar velocity data will provide an additional valuable probe for our approach, since they can constrain $\mu$ directly~\cite{song}, thus breaking the $\Sigma$-$\mu$ degeneracy.

~

\acknowledgments
We are grateful to Liping Fu, Martin Kilbinger and Yannick Mellier for permission to use the CFHTLS data in this paper, and to Catherine Heymans, Ludovic Van Waerbeke and Ismael Tereno for valuable comments on an earlier version of the draft. We also thank Robert Crittenden for useful discussions. The computations were performed on the Western Canada Research Grid (WestGrid) facility. TG acknowledges support from the Alexander von Humboldt Foundation. LP is supported by an NSERC Discovery Grant. AS is supported by the grant NSF AST-0708501. GZ, DB, KK, RCN and YSS are supported by STFC. DB and KK are also supported by RCUK. KK acknowledges support from the European Research Council.

\bibliographystyle{apj}

\end{document}